\documentclass[sigconf]{acmart}
\usepackage{xcolor}
\usepackage{enumitem}
\usepackage{float}
\usepackage[font=small]{caption}

\usepackage[font=small,labelfont=bf]{caption}
\usepackage{paralist}
\usepackage{subfigure,epstopdf}

\AtBeginDocument{
  \providecommand\BibTeX{{
    \normalfont B\kern-0.5em{\scshape i\kern-0.25em b}\kern-0.8em\TeX}}}

\copyrightyear{2020}
\acmYear{2020}
\setcopyright{none}\acmConference[SIGIR '20]{Proceedings of the 43rd International ACM SIGIR Conference on Research and Development in Information Retrieval}{July 25--30, 2020}{Virtual Event, China}
\acmBooktitle{Proceedings of the 43rd International ACM SIGIR Conference on Research and Development in Information Retrieval (SIGIR '20), July 25--30, 2020, Virtual Event, China}
\acmPrice{}
\acmDOI{}
\acmISBN{}



\begin{document}

\title[Reputation (In)dependence in Ranking Systems]{Reputation (In)dependence in Ranking Systems: \\Demographics Influence Over Output Disparities}

\author{Guilherme Ramos}
\orcid{0000-0001-6104-8444}
\affiliation{
  \institution{Dept. of Electrical and Computer Engineering}
  \institution{Faculty of Engineering, University of Porto, Portugal}
}
\email{guilhermeramos21@gmail.com}

\author{Ludovico Boratto}
\orcid{0000-0002-6053-3015}
\affiliation{
  \institution{Data Science and Big Data Analytics}
  \institution{EURECAT, Centre Tecnol\'ogic de Catalunya}
  \city{Barcelona}
  \country{Spain}
}
\email{ludovico.boratto@acm.org}

\renewcommand{\shortauthors}{Guilherme Ramos and Ludovico Boratto}

\begin{abstract}
Recent literature on ranking systems (RS) has considered users' exposure when they are the object of the ranking. Although items are the object of {\em reputation-based RS}, users have a central role also in this class of algorithms. Indeed, when ranking the items, user preferences are weighted by how relevant this user is in the platform (i.e., their reputation). In this paper, we formulate the concept of {\em disparate reputation (DR)} and study if users characterized by sensitive attributes systematically get a lower reputation, leading to a final ranking that reflects less their preferences. We consider two demographic attributes, i.e., gender and age, and show that DR systematically occurs. Then, we propose mitigation, which ensures that reputation is independent of the users' sensitive attributes. Experiments on real-world data show that our approach can overcome DR and also improve ranking effectiveness. 
\end{abstract}

\maketitle

\section{Introduction}\label{sec:intro}
Ranking algorithms are one of our primary forms of interaction with Web content, from search to recommendations. 
The fact that human beings are inherently part of the ranking process has become a topic of prime relevance. 
On the one hand, it is known that the users perceive highly ranked results as more reliable. Hence, a biased ranking would lead to a loss of trust in the system~\cite{PanHJLGG07, RamosBC20}. 
On the other hand, users can also be the object of the ranking (e.g., when dealing with job candidates in services such as LinkedIn), so the ranking position gives them a certain {\em exposure}~\cite{SinghJ18}; if user exposure is affected by their sensitive attributes (such as gender), this might lead to undesired effects, such as discrimination~\cite{HajianBC16}. 

Item rankings are based on user preferences, and the literature has studied the impact of biased rankings on the users~\cite{KulshresthaEMZG19}.
To the best of our knowledge, the impact of users' sensitive attributes in the scope of item ranking systems is an underexplored area. No study analyzes how the ranking systems' underlying mechanisms might lead to a biased ranking w.r.t. users' sensitive attributes.

To tackle this issue in-depth, we focus on a class of ranking systems where each user is given a different relevance by the ranking algorithm, based on a notion of {\em reputation}. 
Specifically, {\em reputation-based ranking systems} score the items by weighting user preferences with the reputation of each user in the system. 
Reputation may be automatically computed based on user behavior or notions of trust~\cite{medo2010effect,li2012robust,saude2017robust}. 
These systems are a form of non-personalized ranking, useful when users are not logged in (e.g., movie rankings in IMDB) or to preserve the system in case of attacks~\cite{li2012robust,saude2017robust,saude2017reputation}. 

\citeauthor{KamishimaAAS18}~\cite{KamishimaAAS18} recently introduced the concept of {\em recommendation independence}. Given a sensitive feature (either associated with the consumers, the providers, or the items), they present a framework to generate recommendations whose outcome is statistically independent of a specified sensitive feature. In this paper, we embrace a similar concept in the ranking systems domain. We propose a method to ensure that, for a given sensitive users' feature, the reputation scores are not biased; indeed, the average of the opinions have the same importance. However, since no personalization exists in our class of ranking systems, our formulation drastically differs from that of \citeauthor{KamishimaAAS18}, and no comparison is possible.

To characterize our problem, we divide users into classes based on the demographic attributes that define them and introduce the concept of {\em disparate reputation (DR)}, capturing if users belonging to different classes are given systematically lower/higher reputation values. 
In our case, we compute reputation solely considering user preferences. 
If DR occurs, then the ranking exhibits a bias based on users' sensitive attributes, thus not reflecting their preferences. 
This bias might lead to negative consequences, such as 
($i$) unfairness towards the consumers, as users belonging to minorities, might receive systematically worse results, 
($ii$) unfairness towards providers whose items are targeted mainly to/preferred by minorities, as these items would systematically get a worse exposure, and ($iii$) trust issues for the platform as a whole, with underwhelmed minorities and providers possibly leaving the platform.

In this work, we show that DR systematically affects the generated rankings, considering two different attributes of the users (gender and age).  
This allows us to study and tackle the problem under binary and multiclass settings\footnote{While gender is by no means a binary construct, to the best of our knowledge, no real-world dataset containing non-binary genders exists. With a binary setting, we mean that we consider a binary gender feature. By also dealing with the age attribute, we show that our problem and solution can be adapted, as is, also to no-binary genders.}. 
To avoid this phenomenon, we propose an algorithm that ensures that reputation is independent of users' sensitive attributes. 
Further, the proposed additional step to introduce reputation independence may be included in any ranking system that computes rankings as a weighted average of the ratings. 
Results of real-world data show the effectiveness of our approach at generating rankings not affected by DR. Besides, thanks to reputation independence, the generated rankings are closer to the primary user preferences, w.r.t. those from state-of-the-art solutions.

Our contributions can be summarized as follows: ($i$) we propose a metric to characterize DR based on users' sensitive attributes, ($ii$) we present an algorithm to introduce reputation independence from sensitive attributes, ($iii$) we measure DR and the effects of our mitigation on real-world data. 

\section{Preliminaries}\label{sub:notation}
\noindent {\bf Notation.} We denote sets by calligraphic letters, e.g., $\mathcal A$, $\mathcal U$ and $\mathcal I$. 
We denote a set of $n\in\mathbb N$ users by $\mathcal U=\{u_1,\ldots,u_n\}$ and a set of $m\in\mathbb N$ items by $\mathcal I=\{i_1,\ldots,i_m\}$. We denote a possibly sparse matrix of ratings that users in $\mathcal{U}$ give to items in $\mathcal{I}$ by $R\in\mathbb R^{n\times m}$. We assume the ratings to be normalized (dividing by the maximum allowed rating) to be in $]0,1]$, and we denote by $\Delta_R$ the difference between the maximum and the minimum normalized ratings. Hence, for $u\in\mathcal{U}$ and $i\in\mathcal{I}$, $R_{ui}=0$ if user $u$ did not rate item $i$, and it is positive otherwise. We denote a set of user attributes by $\mathcal A$, that correspond, for example, to the gender, age, etc. 
An attribute $a\in\mathcal A$ has different classes. For instance, the attribute $a$ that is gender has two or more classes $a=\{male,female, ...\}$. 
We denote classes of an attribute $a\in\mathcal A$ by $a',b',a_1,\ldots,a_k$. 
If user $u\in\mathcal U$ belongs to class $a'\in a$ for attribute $a\in\mathcal A$, then we write that $a(u)=a'$.   
We denote the set of users that rated item $i\in\mathcal I$ by $\mathcal U_i=\{u\in\mathcal U\,:\,R_{ui}>0\}$, the set of items that user $u\in\mathcal U$ rated by $\mathcal I_u=\{i\in\mathcal I\,:\,R_{ui}>0\}$ and the set of users with class $a'\in a$ of attribute $a\in\mathcal A$ by $\mathcal U_{a'}=\{u\in\mathcal U\,:\,a(u)=a'\}$. In this work, we assume that if an attribute $a\in\mathcal A$ has classes $a=\{a_1,\ldots,a_k\}$, then $\mathcal U_{a_i}\cap\mathcal U_{a_j}=\emptyset$ for all $1\leq i,j\leq k$. 
Given a vector $v\in\mathbb R^n$, we denote its \emph{average} by $avg(v)=\frac{1}{n}\sum_{i=1}^n v_i$ and its \emph{standard deviation} by $std(v)=\sqrt{\frac{1}{n}\sum_{i=1}^n \left(v_i-avg(v)\right)^2}$. 

\vspace{0.2cm}
\noindent{\bf Problem Statement.}
Given a dataset with users $\mathcal U$, items $\mathcal I$, ratings given from users to items $R$, and an attribute $a\in\mathcal A$ such that $a=\{a_1,\ldots,a_k\}$, we would like to achieve the following: 

\noindent  $\,(i)\,$  compute users' reputation $\{c_u\}_{u\in\mathcal U}$ based on user preferences;  reputation should capture how revelant are the preferences of a user for the rest of the community, thus excluding the trivial reputation assignment that yields equal reputation for each user;

\noindent $\,(ii)\,$ compute the items' rankings $\{r_i\}_{i\in\mathcal I}$ as a weighted average of users' reputations with items' ratings;

\noindent $\,(iii)\,$ obtain reputations' distributions for each pair of users sets $\mathcal U_{a'}$ and $\mathcal U_{b'}$, where $a',b'\in a\in\mathcal A$ are classes of the same attribute, are statistically indistinguishable ({\em reputation independence}).   

\section{Reputation-based Ranking Systems}\label{sec:rep}

In~\cite{li2012robust}, ~\citeauthor{li2012robust} proposed a reputation-based RS that is an iterative scheme that converges with exponential rate and is more robust to attacks than the arithmetic average (AA).  
At each iteration, the system: (i) estimates the ranking, $r_i^{k+1}$,  of each item $i$ by combining the ratings given to the item with the reputations of users, $c_u^{k}$, that rated the item; (ii) estimates the users' reputation by measuring how different are the user's ratings to the items' ranking estimated in (i). 
Specifically, for the variant L1-AVG, for $k>0$, 
\begin{equation}\label{eq:li}
    \begin{cases}
    r_i^{k+1} = \displaystyle\frac{1}{|\mathcal U_i|}\sum_{u\in \mathcal U}R_{ui} c_u^k\\
    c_u^{k+1} = 1 - \displaystyle\frac{\lambda}{|\mathcal I_u|}\sum_{i\in \mathcal I_u}|R_{ui}-r_i^{k+1}| 
    \end{cases}
\end{equation}
for any initial $c_u^0\in]0,1]$ (we opt to set $c_u^0=1$) and  $\lambda\in]0,1]$ is a hyper-parameter that penalizes the disagreement of ratings with rankings for each user. We denote by $c$ the vector collecting users' reputations in the same order as $\mathcal U$.  
However, this RS presents some unintuitive properties. The ranking is a weighted sum divided by the number of parcels in the summation, instead of being normalized by the sum of the weights. 
Hence, an item such that all the given ratings are maximum will yield a ranking that is not the maximum as long as at least one of the users that rated the item has reputation below 1. 
Therefore, in~\cite{saude2017robust}, the system in~\eqref{eq:li} was enhanced, adjusting the ranking computation of~\eqref{eq:li} to be 
\begin{equation}\label{eq:ramos}
    r_i^{k+1} = \displaystyle\sum_{u\in \mathcal U}R_{ui} c_u^k\big/\displaystyle\sum_{u\in \mathcal U}c_u^k
\end{equation}
This RS also converges with exponential rate and is more robust to attacks than~\eqref{eq:li}. 

\section{Reputation Independence}\label{sec:fair}
This section presents our metric to characterize DR and our mitigation algorithm to ensure  reputation independence from sensitive attributes of the users.

\vspace{0.2cm}
\noindent{{\bf Characterizing disparate reputation (DR)}.} Let $a\in\mathcal A$ be an attribute of the users, with $k>1$ classes $a=\{a_1,\ldots,a_k\}$. Considering two classes $a',b'\in a$ (in the same attribute), we denote as $\mu_{a'}=avg(\{c_u\}_{u\in \mathcal U_{a'}})$ and $\mu_{b'}=avg(\{c_u\}_{u\in \mathcal U_{b'}})$ the average reputation of the users characterized by that class.

We are interested in studying whether a common attribute characterizes polarized reputations for groups of users. 
For example, whether there is a bias in the reputation-based RS such that the opinion of users coming from different classes (e.g., males and females) contribute differently to the ranking computation. 

To characterize if users belonging to different classes have different reputations scores, we define the \textbf{disparate reputation} metric, computed as
   $ \Delta(c)=\mathcal \mu_{a'}-\mu_{b'}.$
The metric ranges in $[-1+\Delta_R\lambda,1-\Delta_R\lambda]$; it is $0$ when both averages of the reputations are the same ($\mu_a=\mu_b$). Negative values point that class $b$ has users with  higher reputation values and, vice-versa, for the class $a$ and positive values. 

To characterize if disparate reputation systematically affects the users belonging to a class, we propose to do a statistical test, the \emph{Mann-Whitney (MW) test}~\cite{mann1947test}, to each pair in $a\in\mathcal A$. It is a nonparametric test,  with the null hypothesis that it is equally likely that a randomly selected value from one population will be less than or greater than a randomly selected value from another population. This test is often used to scrutinize whether two independent samples were selected from populations with the same distribution. 

\vspace{0.2cm}
\noindent{\bf Reputation independence.} To avoid sensitive attributes of the users to impact the ranking system systematically, we design a strategy that, given a sensitive attribute of the users in the system, mitigates the bias in the user reputations for each group of users with different values for that attribute, thus leading the reputation computation to be independent of the sensitive attribute.

Given a reputation-based RS that updates the items' ranking as a weighted average of ratings with users' reputations (such as~\eqref{eq:li} and~\eqref{eq:ramos}), we propose to harmonize users' reputations inside each group of a specific attribute, to achieve a similar distribution of reputations among each group. 
If we consider~\eqref{eq:li} or~\eqref{eq:ramos} to compute rankings and reputations doing $N$ iterations, we use $c_u^N$ and $r_i^N$ to do the following additional step to ensure independence for attribute $a\in\mathcal A$, with classes $ a=\{a_1,\ldots,a_k\}$,
\begin{equation}\label{eq:fair}
    \begin{cases}
    c_u = \mu +\displaystyle \left(c_u^N-\mu_l\right)\frac{\sigma}{\sigma_l}, & \text{for }l=1,\ldots,k\textit{ and }u\in \mathcal U_{a_l}\\
    r_i = \displaystyle\sum_{u\in \mathcal U}R_{ui} c_u\big/\displaystyle\sum_{u\in \mathcal U}c_u &
    \end{cases}
\end{equation}
where $\mu = \min\left\{\mu_1\ldots,\mu_k\right\}$ and $\sigma = \min\left\{\sigma_1\ldots,\sigma_k\right\}$, with\\ $\mu_l=avg\left(\left\{c_u^N\right\}_{u\in U_{a_1}}\right)\text{ and  }\sigma_l=std\left(\left\{c_u^N\right\}_{u\in U_{a_1}}\right).$
Observe that, in Equation~\eqref{eq:fair}, we select the minimum between the averages and the minimum between the standard deviations to ensure that the reputations' readjustment still lies in the interval $]0,1]$. So, Equation~\eqref{eq:fair} harmonizes the reputation's distributions for each class of an attribute to follow a common probability distribution, ensuring that reputations of each class are ``statistically indistinguishable''. 

\section{Evaluation on Real Data}\label{sec:ill}
Capturing and dealing with DR is not a trivial task due to the lack of public datasets with ratings and sensitive attributes of the users. This led us to this preliminary study, whose goal is to illustrate the problem and to validate it considering different sensitive attributes. 

Here, we compare the state-of-the-art RS proposed in~\cite{saude2017robust} and computed as in~\eqref{eq:ramos}, and the solution we introduce in~\eqref{eq:fair}. We do this both in terms of DR and ranking effectiveness.

We use the \emph{MovieLens-1M} dataset, which has 1 000 209 ratings from $|\mathcal U|=$ 6 040 users to $|\mathcal I|=$ 3 952 items. We evaluate our work on users' attributes $\mathcal A=\{gender,age\}$, available in the dataset. 

\subsection{Evaluating Disparate Reputation}

\begin{figure}
\centering      
\includegraphics[width=0.27\textwidth]{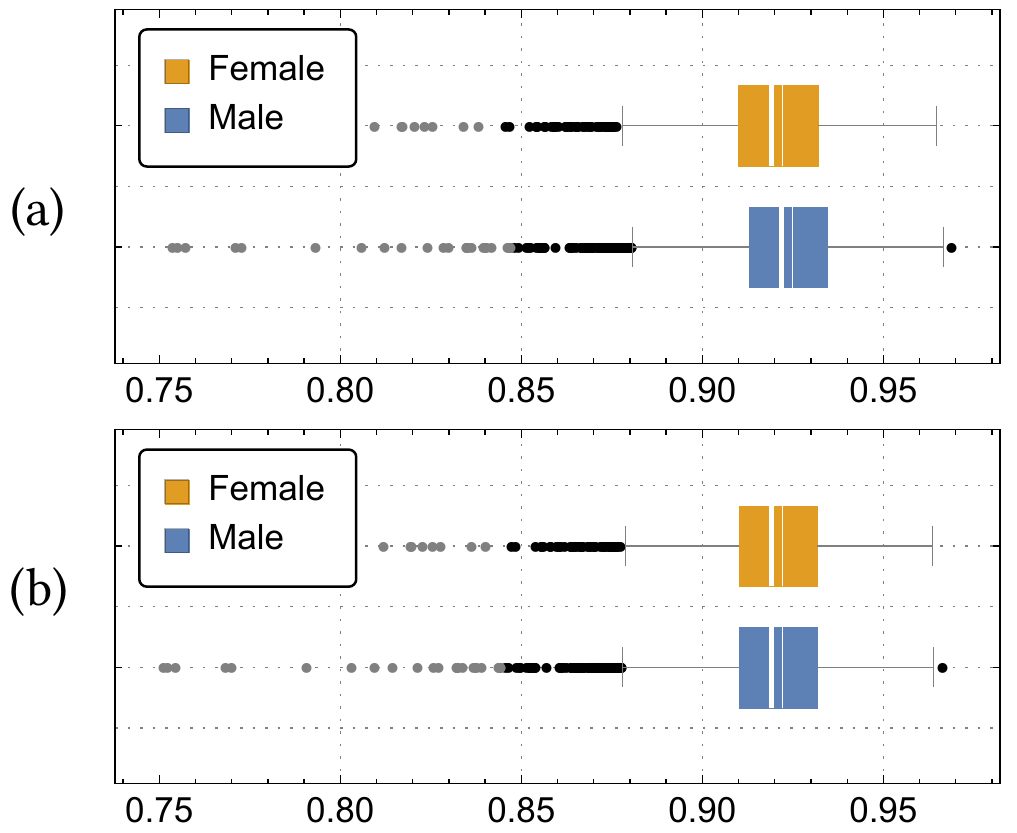}
\caption{BWC for reputations of users resulting from~\eqref{eq:ramos} in (a), and from ~\eqref{eq:ramos} and~\eqref{eq:fair} in (b), with $\lambda = 0.5$, for groups of users under the attribute gender.}
\label{fig:genderBWC} 
\end{figure}

\noindent{\bf Attribute: Gender.}
First, we investigate if there is bias on users' reputations under the attribute gender. 
We start by characterizing DR by presenting 
the Box-whisker-chart (BWC) for the reputations, see \figurename~\ref{fig:genderBWC} (a).  
Using the proposed DR metric to assess bias concerning the attribute gender. 
Using solely equation~\eqref{eq:ramos}, we obtain $\Delta(c)=\mu_\text{female}-\mu_\text{male}\approx -0.0027$. Hence, since $\Delta(c)\in[-1+0.8\times0.5,1-0.8\times0.5]=[-0.6,0.6]$, it follows that the class of male users do have in average larger reputation values than the class of female users, yielding a bias on the attribute gender (as we observe also in \figurename~\ref{fig:genderBWC} (a)). Next, we test the null hypothesis that the median difference is 0 at the 5\% level based on the MW test.
The hypothesis is rejected with a p-value of $8.240\times 10^{-8}$. 
Hence, we confirm that there is bias in the reputations for these two classes. 

To mitigate the bias, we compute the reputation and the ranking as presented in equation~\eqref{eq:fair}, obtaining 
the BWC for reputations of \figurename~\ref{fig:genderBWC} (b). 
In this case, we get a DR of $\Delta(c)\approx -2.2205\times 10^{-16}\approx 0$. Hence, we mitigate the bias on the attribute gender (as we observe also in \figurename~\ref{fig:genderBWC} (b)).
This time, the null hypothesis that the median difference is 0 is not rejected at the 5\% level based on the MW test, with a p-value of $0.9505$. 
This result confirms that we successfully mitigated the bias on the reputations for these two classes.   

\begin{table}
\caption{MW tests for the reputations resulting from~\eqref{eq:ramos} of each pair of classes. $H_0$ means $H_0$ is not rejected and $H_1$ means $H_0$ is rejected.}

$\resizebox{.472\textwidth}{!}{
\begin{tabular}{c|ccccccc}
 & $(<18)$ & $(18-24)$ & $(25-34)$ & $(35-44)$ & $(45-49)$ & $(50-55)$ & ($>55$)
\\ \hline
$(<18)$ & -- & $H_1$ & $H_1$ & $H_1$ & $H_1$ & $H_1$ & $H_1$ \\
$(18-24)$ & -- & -- & $H_1$ & $H_1$ & $H_1$ & $H_1$ & $H_1$ \\
$(25-34)$ & -- & -- & -- & $H_1$ & $H_0$ & $H_0$ & $H_1$ \\
$(35-44)$ & -- & -- & -- & -- & $H_0$ & $H_0$ & $H_0$ \\
$(45-49)$ & -- & -- & -- & -- & -- & $H_0$ & $H_0$ \\
$(50-55)$ & -- & -- & -- & -- & -- & -- & $H_0$ \\
\end{tabular}
}$
\label{tab:1}
\end{table}
\begin{table}
\caption{DR of reputations resulting from~\eqref{eq:ramos}, for attribute age.}
$\resizebox{.472\textwidth}{!}{
\begin{tabular}{c|ccccccc}
 & $(<18)$ & $(18-24)$ & $(25-34)$ & $(35-44)$ & $(45-49)$ & $(50-55)$ & ($>55$)
\\ \hline
$(<18)$ &  -- & -0.0089 & -0.0142 & -0.0161 & -0.0159 & -0.0153 & -0.0164 \\
$(18-24)$ & -- & -- & -0.0053 & -0.0072 & -0.0070 & -0.0064 & -0.0075 \\
$(25-34)$ & -- & -- & -- & -0.0019 & -0.0017 & -0.0011 & -0.0022 \\
$(35-44)$ & -- & --& -- & -- & 0.0002 & 0.0008 & -0.0003 \\
$(45-49)$ & -- & -- & -- & -- & -- & 0.0006 & -0.0005 \\
$(50-55)$ & -- & -- & -- & -- & -- & -- & -0.0011 \\
\end{tabular}
}$
\label{tab:3}
\end{table}

\begin{figure}
\centering     
\includegraphics[width=0.27\textwidth]{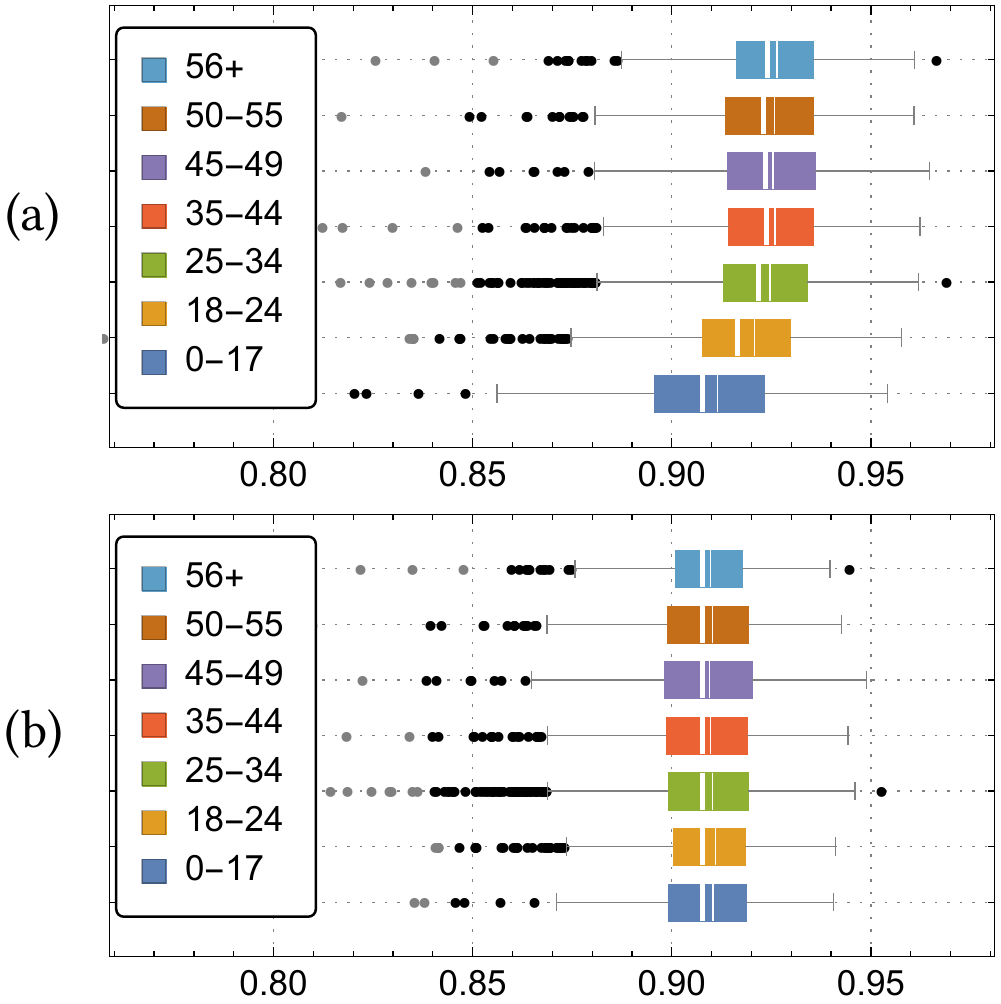}
\caption{BWC for reputations of users resulting from~\eqref{eq:ramos} in (a), and from ~\eqref{eq:ramos} and~\eqref{eq:fair} in (b), with $\lambda = 0.5$, for groups of users under the attribute age.}
\label{fig:ageBWC} 
\end{figure}

\vspace{0.2cm}
\noindent{\bf Attribute: Age.} We characterize if DR occurs, by computing the Box-whisker-chart (BWC) for the reputations, see \figurename~\ref{fig:ageBWC} (a). 

Subsequently, we perform a similar analysis to the attribute age, where we consider groups of users with age in the classes $\{(<18) , (18-24) , (25-34) , (35-44) , (45-49) , (50-55) , (>55)\}$. 

The DR metric, when only~\eqref{eq:ramos} is used, yields the results in Table~\ref{tab:3} that reveal the existence of bias. We only filled the up-triangular part of the table, because the DR anti-commutes and the low-triangular part is equal to the symmetric of the up-triangular one.  
The result of the MW test for the reputations, resulting from~\eqref{eq:ramos} of each pair of classes, for the null hypothesis that the median difference is 0 ($H_0$) at a 5\% confidence level vs. the hypothesis that the median difference is not 0 ($H_1$) is summarized in Table~\ref{tab:1}. 
We only filled the up-triangular part of the table since the MW test commutes. 

When we mitigate bias for the attribute age with~\eqref{eq:fair}, we achieve the results in Table~\ref{tab:4} and Fig.~\ref{fig:ageBWC} (b).  
Now, the null hypothesis that the reputations' median difference is 0 ($H_0$) is not rejected at the 5\% confidence level, using the MW test, for any pair of classes under attribute age (table is not reported due to space constraints). 

\begin{table}
\caption{DR of reputations resulting from~\eqref{eq:ramos} and~\eqref{eq:fair}, for attribute age.}
$\resizebox{.472\textwidth}{!}{
\begin{tabular}{c|ccccccc}
 & $(<18)$ & $(18-24)$ & $(25-34)$ & $(35-44)$ & $(45-49)$ & $(50-55)$ & ($>55$)
\\ \hline
 $(<18)$ & -- & $2\times 10^{-16}$ & $1\times 10^{-16}$ & $-2\times 10^{-16}$ & $4\times 10^{-16}$ & $2\times 10^{-16}$ & $1\times 10^{-16}$ \\
 $(18-24)$ & -- & -- & $-1\times 10^{-16}$ & $-4\times 10^{-16}$ & $2\times 10^{-16}$ & 0 & $-1\times 10^{-16}$ \\
 $(25-34)$ & -- & -- & -- & $-3\times 10^{-16}$ & $3\times 10^{-16}$ & $1\times 10^{-16}$ & 0 \\
 $(35-44)$ & -- & -- & -- & -- & $6\times 10^{-16}$ & $4\times 10^{-16}$ & $3\times 10^{-6}$ \\
 $(45-49)$ & -- & -- & -- & -- & -- & $2\times 10^{-16}$ & $-3\times 10^{-16}$ \\
 $(50-55)$ & -- & -- & -- & -- & -- & -- & $-1\times 10^{-16}$ \\
\end{tabular}
}$
\label{tab:4}
\end{table}
\subsection{Evaluating Effectiveness}

\begin{table}
\caption{$\tau$ using AA rankings as the ground truth and rankings obtained with~\eqref{eq:ramos}.}
{\scriptsize
\begin{tabular}{ccc}
   No attribute & Gender & Age 
\\ \hline
 $\tau=0.9950$ & $\tau=0.9954$ & $\tau=0.9959$ 
 \\
\end{tabular}
\label{tab:5}
}
\end{table}
Finally, to evaluate the effectiveness of the proposed method, we use the Kendall Tau~\cite{kendall1938new} with AA as the ground truth, as it is done in~\cite{li2012robust}. 
We report the observed $\tau$ for each of the attributes considered in Table~\ref{tab:5}. We notice that the Kendall Tau improves when we mitigate bias relative to an attribute. This improvement means that our approach yields an order of rankings closer to the AA, but yet assigning different relevance to different users, w.r.t.~\eqref{eq:ramos}.  

The reputation concept treats users differently, which may lead to a ranking with a bias for specific users' attributes. 
With the proposed approach, for a specific attribute, we mitigate bias. 
With our method, the concept of reputation still plays a role inside each group with a particular attribute value, but it does not cause bias. 
So, we get ``closer'' to the average in the sense that the AA does not treat groups differently, and with our approach, we also do not treat groups differently for a given attribute. 

\section{Conclusions}\label{sec:conc}
Reputation-based ranking systems try to rank items by ensuring the preferences of the community as a whole are reflected in the way items are sorted. In this sense, computing effective formulations of {\em user reputation}, to weight the individual preferences, is vital. 

In this work, we introduce a measure of {\em disparate reputation (DR)}, to analyze if user reputation is affected by users' sensitive attributes. To avoid this, we introduce a novel approach that ensures reputation independence from sensitive user attributes. Experiments on real data, which considered different demographic attributes of the users, showed that DR occurs in state-of-the-art approaches and that our mitigation can introduce reputation independence from sensitive attributes and, at the same time, increase ranking quality.

Avenues for further research include exploring further datasets, considering attributes with possibly not disjoint classes, and specifying multiple attributes to mitigate bias. 

{\small
\noindent\textbf{Acknowledgments.} This work was supported in part by FCT project POCI-01-0145-FEDER-031411-HARMONY. 
G. Ramos further acknowledges the support of Institute for Systems and Robotics, Instituto Superior T\'ecnico (Portugal), through scholarship BL229/2018\_IST-ID.
L. Boratto acknowledges Ag\`encia per a la Competivitat de l'Empresa, ACCI\'O, for their support under project ``Fair and Explainable Artificial Intelligence (FX-AI)''. 
}
\bibliographystyle{ACM-Reference-Format}
\bibliography{sample-base}

\end{document}